\documentclass[11pt,preprint,showpacs]{revtex4}
\usepackage{graphicx}
\usepackage{dcolumn}
\usepackage{bm}
\usepackage{amsmath}
\usepackage{amssymb}
\usepackage{amsfonts}

\begin{document}

\title{Proton-proton scattering without Coulomb force renormalization}

\author{R.~Skibi\'nski}
\affiliation{M. Smoluchowski Institute of Physics, Jagiellonian
University,
                    PL-30059 Krak\'ow, Poland}
\author{J.~Golak}
\affiliation{M. Smoluchowski Institute of Physics, Jagiellonian
University,
                    PL-30059 Krak\'ow, Poland}
\author{H.~Wita{\l}a}
\affiliation{M. Smoluchowski Institute of Physics, Jagiellonian
University,
                    PL-30059 Krak\'ow, Poland}
\author{W.\ Gl\"ockle}
\affiliation{Institut f\"ur theoretische Physik II,
Ruhr-Universit\"at Bochum, D-44780 Bochum, Germany}

\date{\today}

\begin{abstract}
We demonstrate numerically that proton-proton (pp) scattering observables
can be determined directly by standard short
range methods using a screened pp Coulomb force without renormalization.
 In examples  the
appropriate screening radii are given. We also numerically investigate
solutions of the 3-dimensional Lippmann-Schwinger (LS)
 equation for a screened Coulomb potential alone in the limit of large
screening radii and confirm analytically predicted properties for off-shell,
half-shell and on-shell Coulomb  t-matrices. These 3-dimensional solutions
will form a basis for a novel approach to include the pp Coulomb interaction
into the
 3N Faddeev framework. \end{abstract}


\maketitle \setcounter{page}{1}

\section{Introduction}

The action of the Coulomb force in pp scattering can be rigorously treated
using the Vincent-Phatak method~\cite{Vincent74}.
We propose an alternative manner using a screened Coulomb force, despite the
well know fact  that the screening limit does not exist. Namely the pp
on-shell scattering amplitude acquires an oscillating phase factor if the
screening radius goes to infinity~\cite{chen72,taylor1,taylor2,Alt78}.
 This phase factor is known  and can be
removed, a step known in that context under the name renormalization.
However, as we shall show, if one is
interested in the pp observables (not in the phase shifts) where that phase
factor drops out, all
scattering observables can  be obtained in the  standard framework of short
range interactions. This will be
demonstrated in section  II for suitably chosen screening radii.

In view of a forthcoming paper~\cite{Witala08} related to the pd system, we
further investigate
 in section III properties of the screened 3-dimensional Coulomb t-matrix $<
\vec p~' | t_c^ R(E) | \vec p> $. This t-matrix  is a solution of
 the 3-dimensional
2-body LS equation driven by the screened Coulomb potential.
 Namely to catch the full action of the Coulomb force in
the pd system
 a partial wave truncated pp t-matrix is insufficient and the complete
3-dimensional Coulomb
 t-matrix has to be used. Analytical properties of that screened Coulomb
t-matrix,
 off-the-energy-shell, half-shell and  on-shell have been  studied in the
past~\cite{chen72,ford1964,ford1966,taylor1,taylor2}.
These investigations, however, mostly rely on insights gained for fixed
partial wave
states. The mathematical rigor
in the summation of the partial wave sum to infinity leaves room
for improvement. Therefore we
felt that a numerical study is justified to verify statements given there:
 the  screening limit of $ < \vec p~' | t_c^R (E) | \vec p> $  exists for $
\frac{p'^2}{m}\ne E \ne \frac{p^2}{m} $  and coincides with the unscreened
pure Coulomb force expression, which is known
analytically~\cite{chen72,kok1980}
and references therein;
that screening limit  exhibits
a discontinuity if  $ p $ approaches $ \sqrt{m_p E}, E > 0 $ from  above or
below;
the screening limit of the on-shell  t-matrix $ < p \hat p' | t_c^R (E =
\frac{p^2}{m}) | \vec p> $
approaches  the analytically known  unscreened Coulomb on-shell
t-matrix up to a given infinitely oscillating phase factor.
 Here we want to numerically investigate at which R-values these limits are
reached with adequate accuracy.
We conclude in section IV.

\section{The on-shell pp t-matrix with screened Coulomb potential and the pp
observables}

Let $V_c^R$ be the screened Coulomb potential between 2 protons
normalised such  that $V_c^R$ turns into the pure pp Coulomb potential for
$R$, the screening radius, going to infinity. Together with the strong
interaction V
this determines the  2-body pp t-matrix via the LS equation
\begin{eqnarray}
t = V + V_c^R + ( V + V_c^R) G_0 t ~,
\label{LS}
\end{eqnarray}
where  $G_0$ is  the free propagator. That equation is solved at
the $pp$  c.m. energy $E=\frac {p^2} {m_p}$ projected on a set of
partial wave basis states $|p (ls)j m; t m_t>$,  with $p,l,s,j$ and $m$
 the relative momentum, orbital angular momentum,
 total spin, total angular momentum and its magnetic quantum number.

The total isospin quantum numbers for two protons are $t=1$ and $m_t=-1$.
This leads to the on-the-energy-shell t-matrix element
\begin{eqnarray}
< p ( l' s') j' m'| t | p ( ls) j m> = \delta_{s's} \delta_{j'j}
\delta_{m'm}  t_{l'l}^{sj}(p,p) ~,
\end{eqnarray}
where the Pauli principle dictates $(-)^{l+s} = 1$ and we took $s$
to be conserved.

The full 3-dimensional antisymmetrized on-shell t-matrix is given
as
\begin{eqnarray}
< \vec p~' m_1' m_2'  | t ( 1 - P_{12}) | \vec p m_1 m_2 > ~,
\end{eqnarray}
where $m_i (m_i')$  are the individual spin  magnetic quantum
numbers and $\vec p = p\hat p, \vec p~' = p\hat p~'$ the
 initial and final relative momenta.

The standard partial wave decomposition leads to
\begin{eqnarray}
&& < \vec p~' m_1' m_2'  | t ( 1 - P_{12}) | \vec p m_1 m_2 > =
\sum_s  ( \frac{1}{2} \frac{1}{2} s, m_1' m_2' m_s')
   (\frac{1}{2} \frac{1}{2} s, m_1 m_2 m_s) \cr
&& \sum_{j= 0 }^{\infty}\sum_{m = -j}^j \sum_{l'= | j-s|}^{j+s}
 \sum_{l= | j-s|}^{j+s}  \sum_{m_l'} ( l' s j, m_l', m_s',m) Y_{l'm_l'} (
\hat p~')t_{l'l}^{sj}(p,p) ( 1 + ( -)^{l+s} )\cr
&& \sum_{m_l} ( l s j, m_l m_s,m ) Y_{lm_l}^* (\hat p) ~.\label{4}
\end{eqnarray}

The strong force can be neglected beyond a certain $ j_{max}$
and there only the screened Coulomb t-matrix $ t_{cl}^R $ is
present, which is diagonal in $l $  and independent of $s $ and $j$.
In a well known manner one adds and subtracts a finite sum up
to $ j_{max}$ with $t_{cl}^R $ only and this completes the
infinite sum over $j$ containing only $t_{cl}^R $. That infinite sum is
identical to the 3-dimensional antisymmetric screened Coulomb t-matrix.
Thus (\ref{4}) turns into
\begin{eqnarray}
&& < \vec p~' m_1' m_2'  | t ( 1 - P_{12}) | \vec p m_1 m_2 > =
\delta_{m_1' m_1} \delta_{m_2' m_2} < \vec p~'| t_c^R|\vec p > -
\delta_{m_1' m_2} \delta_{m_2' m_1} < \vec p~'| t_c^R|-\vec p >\cr
& + &  \sum_s  ( \frac{1}{2} \frac{1}{2} s, m_1' m_2' m_s')
(\frac{1}{2} \frac{1}{2} s, m_1 m_2 m_s)
 \sum_{j= 0 }^{\infty}\sum_{m = -j}^j \sum_{l'= | j-s|}^{j+s}
\sum_{l= | j-s|}^{j+s} \sum_{m_l'} ( l' s j, m_l', m_s',m) \cr
&& Y_{l'm_l'} ( \hat p~')( t_{l'l}^{sj}(p,p) - \delta_{l'l} t_{cl}^R) ( 1 + (
-)^{l+s} )
\sum_{m_l} ( l s j, m_l m_s,m ) Y_{lm_l}^* (\hat p) ~\label{5}.
\end{eqnarray}

Now as is well known~\cite{taylor1,taylor2}  the limit of that expression
does not
exist for $ R \rightarrow \infty$. In that limit each term in (\ref{5})
acquires the same  infinitely oscillating factor $ e^ {2i \Phi_R(p)}$,
 where $\Phi_R(p)$ is given below.
 If one is interested  in scattering phase shifts it is unavoidable to keep
 track of this oscillating factor which in that context runs under the name
 renormalization~\cite{Alt78}. However, if one is interested in the pp
observables,
the cross  section and all sorts of spin observables (note $\Phi_R(p)$ is
independent
 of spin magnetic quantum numbers), where the on-shell t-matrix appears
 together with its complex conjugate,  the oscillating factor drops out.
 In that case one does not even has to know the analytical form of
$\Phi_R(p)$.
 It is sufficient to know that the limit of large screening radius generates
 just a phase factor.

This is the main message of this section: the pp observables based on the
strong
and the screened Coulomb force  can be  calculated without renormalization
using standard short range
methods. Though not explicitly stated in \cite{taylor1,taylor2} this
insight is in the spirit of these authors.
It remains to establish the values of the parameter $R$ at which
the observables get independent of $R$.

Based on (\ref{5}) all pp-scattering observables are given by
well known analytic expressions~\cite{gloeckle83}, here, however, we use a
more
modern nomenclature for the
various spin observables~\cite{physrep96}. We use the following screening
form  which depends
on two parameters, the screening radius $R$ and the power $n$:
\begin{equation}
V_c^R(r) = \frac{e^ 2} {r} e^{-{(\frac {r} {R})}^n} ~.
\label{eq.2}
\end{equation}

At a given value $n$ the pure Coulomb potential results for $R \rightarrow
\infty$.
 We use $n=1$, $2$, $3$, and $4$. As has been shown in~\cite{kamada05} based
on~\cite{taylor1,taylor2},  the related phase $ \Phi_R(p)$ is given as
\begin{eqnarray}
\Phi_R(p) = -\eta [ ln(2pR) - {\gamma}/n] \label{eq.8}
\end{eqnarray}
where  $\gamma=0.5772\dots$ is the Euler number and $\eta = \frac{ m_p e^
2}{2
p} $ the Sommerfeld
parameter.

Considering only the screened Coulomb force (\ref{eq.2}) the leading term in
 (\ref{LS}) is given by
\begin{eqnarray}
<\vec p~'|V_c^R|\vec p>  =  \frac {e^ 2} {2\pi ^2}\frac{1}{|\vec p - \vec
p~'|}
 \int_0^{\infty} dr sin(|\vec p - \vec p~'|r) e^{-(\frac {r} {R})^n } ~.
\label{eq.c1}
\end{eqnarray}
By a simple partial integration  it can be shown that
  this matrix element approaches  the n-independent limit
\begin{eqnarray}
\lim_{R \to \infty } <\vec p~'|V_c^R|\vec p> = \frac {e^ 2}
 {2\pi ^2|\vec p - \vec p~'|^ 2}
~.
\label{eq.c2}
\end{eqnarray}
Therefore for suitably large $R$-values the leading term and thus  the
solution of the LS equation (\ref{LS}) will approach a n-independent limit.

\begin{figure}
\includegraphics[scale=0.9]{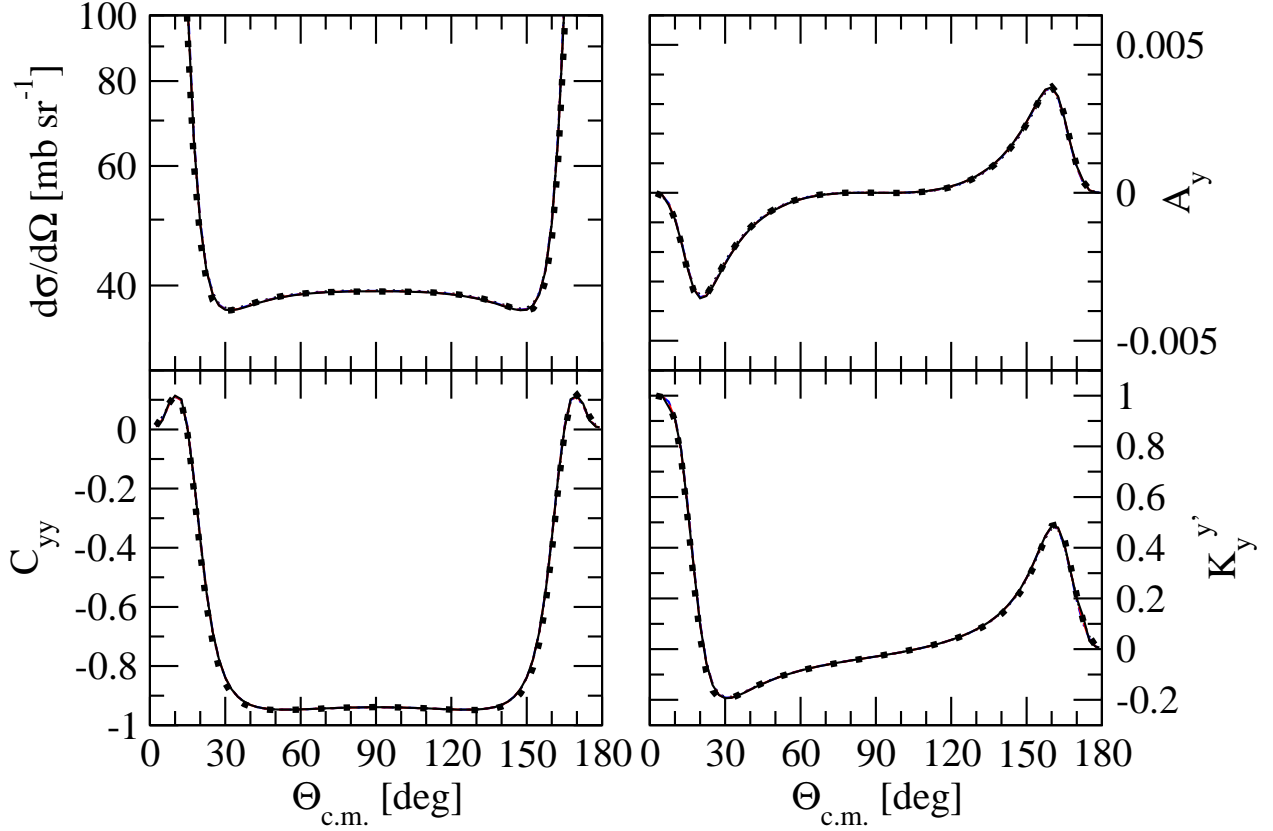}
\caption{(color online) The convergence of the  pp scattering cross section
($\frac {d\sigma} {d\Omega}$), analyzing power ($A_y$), spin correlation
coefficient ($C_{yy}$), and spin transfer coefficient ($K_y^{y'}$) at
$E_p^{lab}=13$~MeV as a
function of the c.m. scattering angle calculated with the screened Coulomb
force and the CD~Bonn nucleon-nucleon potential~\cite{cdbonn}, 
which is  kept for the partial waves up to $j
\le 3$. The screening
 radius  is $R=120$~fm and  $n=1$ (dotted line), $n=2$ (dashed-dotted line),
$n=3$ (dashed line),
 and $n=4$ (solid line). The curves for $ n=1 $ to $ n=4$
all overlap on the scale of the figure. The exact Vincent-Phatak result is
given by thick dots.}
 \label{fig1}
\end{figure}

\begin{figure}
\includegraphics[scale=0.9]{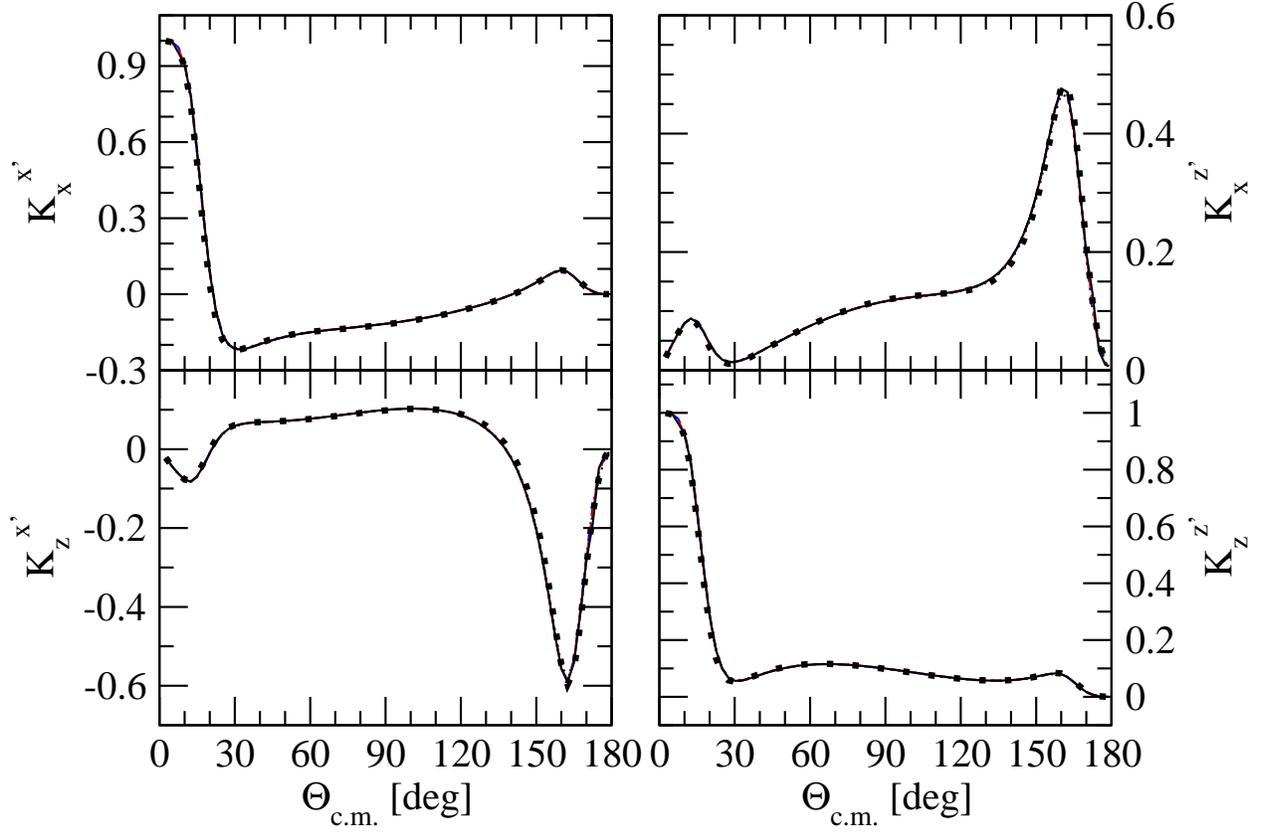}
\caption{(color online) The same as in Fig.\ref{fig1} but for other pp
scattering  spin transfer
coefficients.}
 \label{fig2}
\end{figure}

In Figs.~\ref{fig1} and \ref{fig2}  we  demonstrate at  $E_p^{lab}=13$~MeV
for several pp observables independence on n for a suitably large R-value
and the perfect agreement to the exact Vincent-Phatak~\cite{Vincent74}
results at  $E_p^{lab}=13$~MeV. The  deviation with respect to the
standard Vincent-Phatak  approach~\cite{Vincent74}, which treats the pp
Coulomb
   force rigorously,  for different pp observables and
 values of $n=1, 2, 3$ and $4$ and screening radius $R=120$~fm is under 
$\approx 1 \%$.

\begin{figure}
\includegraphics[scale=0.9]{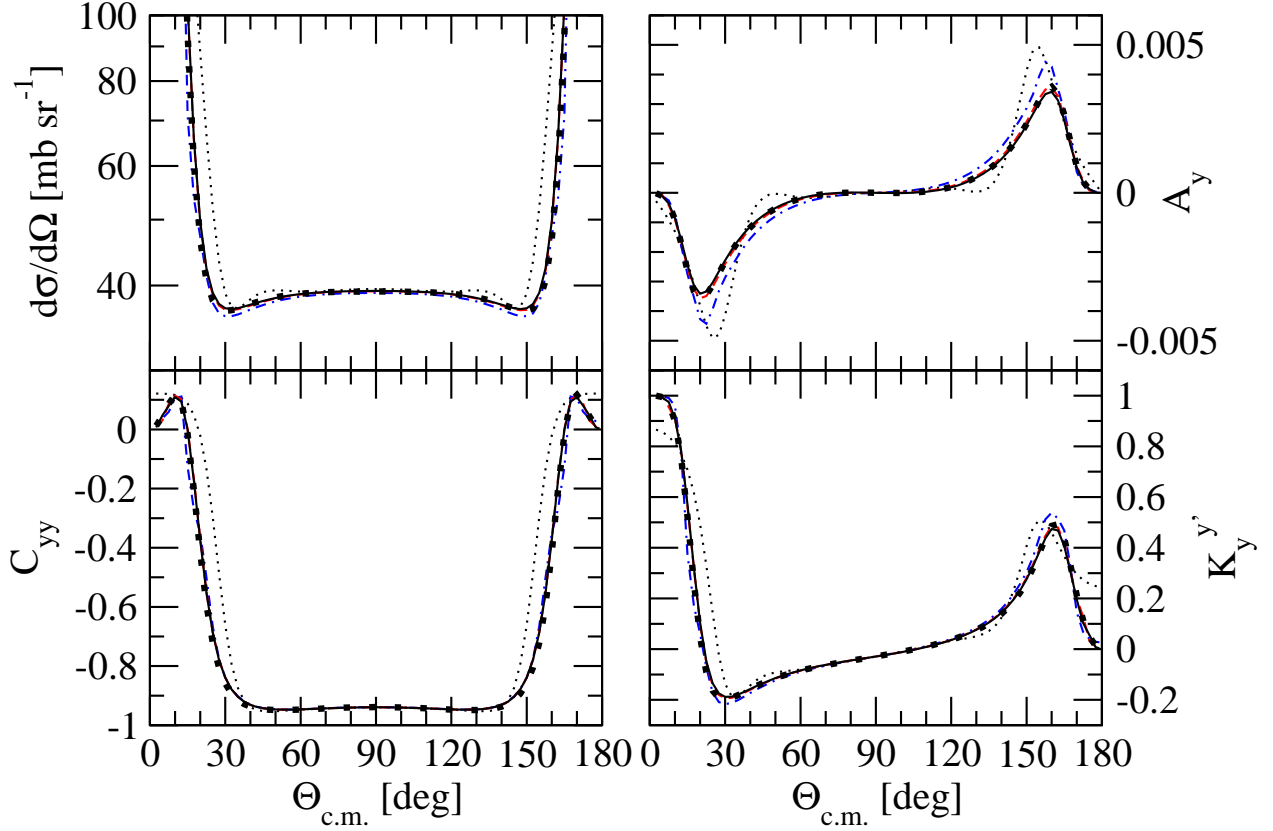}
\caption{(color online) The pp scattering cross section
($\frac {d\sigma} {d\Omega}$), analyzing power ($A_y$), spin correlation
coefficient ($C_{yy}$), and spin transfer coefficient ($K_y^{y'}$) at
$E_p^{lab}=13$~MeV
 as a function of the c.m. scattering angle calculated with the screened
Coulomb
force and the CD~Bonn nucleon-nucleon potential~\cite{cdbonn}, 
which is  kept for the partial waves up to $j
\le 3$.
The screened results are  for  $n=4$  and different values of the screening
radius
 $R$: $R=20$~fm (dotted line), $R=60$~fm (dashed-dotted line),
 $R=120$~fm (dashed line), and $R=180$~fm (solid line).
 The thick dots are the Vincent-Phatak's exact results.}
 \label{fig3}
\end{figure}

\begin{figure}
\includegraphics[scale=0.9]{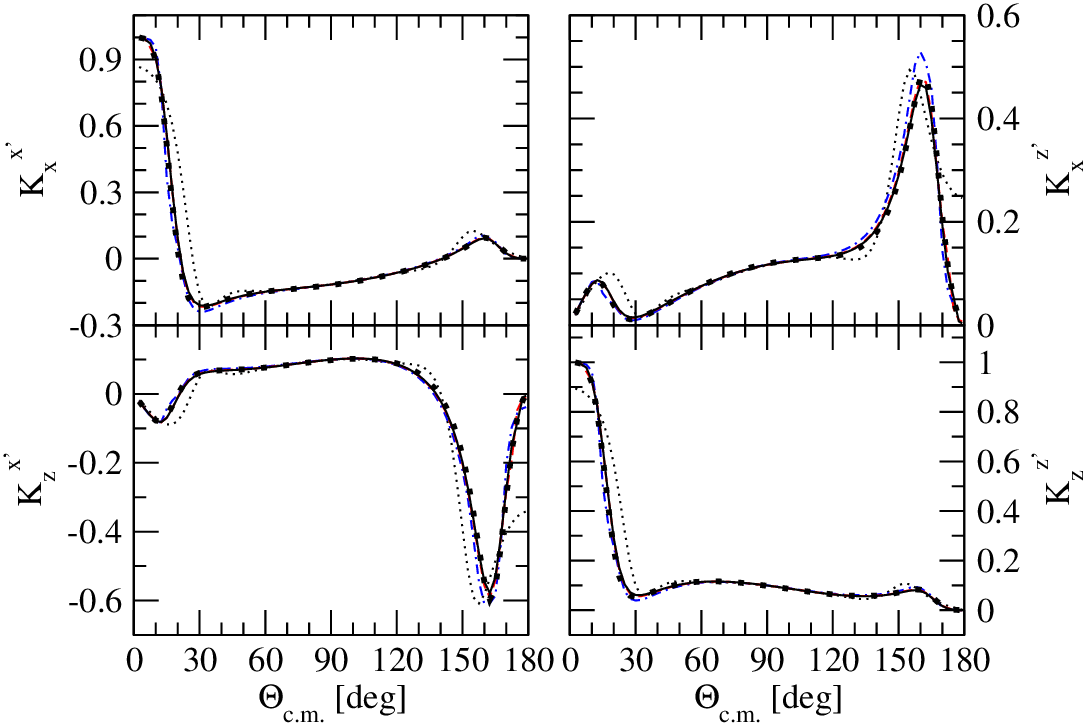}
\caption{(color online) The same as in Fig. \ref{fig3} for other spin
transfer
coefficients.}
 \label{fig4}
\end{figure}

\begin{figure}
\includegraphics[scale=0.9]{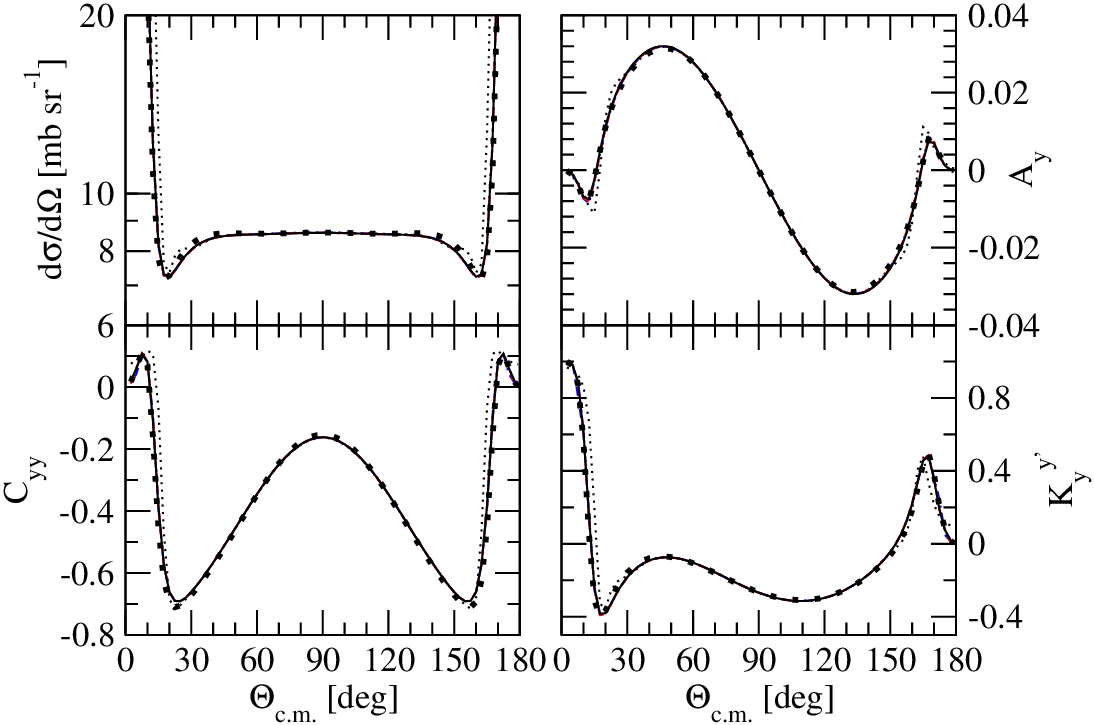}
\caption{(color online) The same as in Fig.\ref{fig3} at $
E_p^{lab}=50$~MeV.}
 \label{fig5}
\end{figure}

\begin{figure}
\includegraphics[scale=0.9]{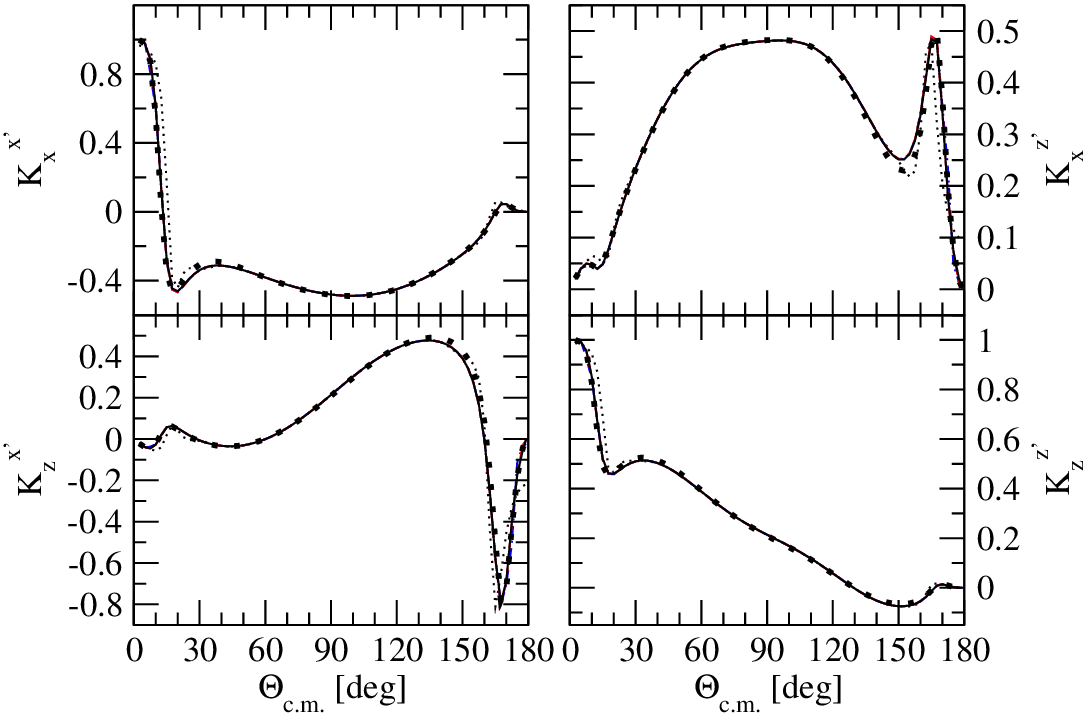}
\caption{(color online) The same as in Fig.\ref{fig4} at
$E_p^{lab}=50$~MeV.}
 \label{fig6}
\end{figure}

In  Figs.~\ref{fig3} and \ref{fig4} we  show  the convergence with
respect to $R$ for $n=4$ for a number of pp observables
 at  $E_p^{lab}=13$~MeV.
 The corresponding results for  $E_p^{lab}=50$~MeV
 are shown in Figs.~\ref{fig5} and \ref{fig6}.
 The resulting limiting values agree very well with the Vincent-Phatak
results.
 At $E_p^{lab}=13$~MeV the limiting $R$-value
is $R=120$~fm and at $E_p^{lab}=50$~MeV $R=60$~fm. With decreasing energy 
 the limiting  $R$-value increases.

\section{Properties of the 3-dimensional screened Coulomb t-matrix}

As will be shown in a forthcoming article~\cite{Witala08} the 3-dimensional
screened pp Coulomb t-matrix
 $ < \vec p~' | t_c^R (E) | \vec p> $ occurs naturally in a certain type of
Faddeev equation of the pd
 scattering problem. There it appears off-the-energy-shell (with the
exception of isolated points). For the unscreened pure  Coulomb force the
off-the-energy-shell expression is analytically known~\cite{chen72,kok1980}.

We numerically investigate the  screening limits of
 $ < \vec p~' | t_c^R (E) | \vec p> $ for $ \frac{p'^2}{m}\ne E \ne
\frac{p^2}{m} $ (off-shell),
for $ p' \ne p$ and $ E= \frac{p^ 2}{m_p} $ (half-shell) and for $ p'= p $
and $ E= \frac{p^ 2}{m_p} $ (on-shell). 
Here we  want to numerically investigate at
which $R$-values
 the limits are reached with adequate accuracy and how they are related to
the corresponding
 unscreened pure Coulomb force expressions.

To that aim we regard the  LS equation for two protons interacting only
with the screened Coulomb potential $V_c^R$.  The off-shell t-matrix element
$< \vec p~' | t_c^R(E = \frac {k^2} {m_p}) | \vec p > \equiv
t_c^R(p',p,x=\hat p \cdot {\hat p}~';E)$ fulfills for given energy E the
equation~\cite{elster1998}
\begin{equation}
t_c^R(p',p,x) = \frac {1} {2\pi} v_c^R(p',p,x,1) +
\int_0^{\infty} dp"p"^2 \int_{-1}^1 dx" v_c^R(p',p",x,x")
\frac{1} {E+i\epsilon - \frac {p"^2} {m_p} } t_c^R(p",p,x") \label{eq.1}
\end{equation}
with
\begin{equation}
v_c^R(p',p,x',x) \equiv \int_0^{2\pi} d\phi
V_c^R(p',p,x'x+\sqrt{1-x'^2} \sqrt{1-x^2} cos\phi) ~.
\label{eq.1a}
\end{equation}

Eq. (\ref{eq.1}) can be solved after discretizing $x$ and the  continuous
momentum variables using direct matrix inversion or generating the
Neumann series  and applying  Pad\`e summation.

We solved (\ref{eq.1}) at three energies: $E=3$~MeV, $13$~MeV, and
$50$~MeV. For $n=1$
the leading term in (\ref{eq.1}) can be calculated analytically
\begin{equation}
v_c^R(p',p,x',x) = \frac {e^2} {\pi \sqrt{(p'^2+p^2-2p'px'x +
\frac {1} {R^2})^2 -4p'^2p^2(1-x'^2)(1-x^2) }} ~. \label{eq.3}
\end{equation}

For $n > 1$ this is no more possible and a two-dimensional
numerical integration is required to get the leading term
\begin{eqnarray}
v(p',p,x',x) = \frac{e^2} {2\pi^2} \int_0^{2\pi} d\phi
\int_0^{\infty} dr \frac
{1}
{\sqrt{p^2+p'^2-2pp'(x'x\sqrt{1-x'^2}\sqrt{1-x^2}cos\phi)}}
e^{-(\frac {r} {R})^n} ~. \label{eq.4}
\end{eqnarray}

On the other hand the  pure off-shell Coulomb t-matrix is  known
analytically (\cite{chen72,kok1980} and references therein)
\begin{eqnarray}
< \vec p~' | t_c^R(\frac {k^2} {m_p}) | \vec p > \rightarrow
\frac{e^2} {2\pi^2} \frac {1+I(x)} {(\vec p~' - \vec
p)^2} \label{eq.7}
\end{eqnarray}
with
\begin{eqnarray}
I(x) = \frac {1} {x}[ _2F_1(1,i\eta;1+i\eta; \frac {x+1} {x-1}) -
  _2F_1(1,i\eta;1+i\eta; \frac {x-1} {x+1})]
\label{eq.7a}
\end{eqnarray}
and  $x^2 = 1 + \frac {(p'^2-k^2)(p^2-k^2)} {k^2(\vec p' - \vec
p)^2} $. $_2F_1$ is the hypergeometric function~\cite{abrom}, which 
we determined using subroutines from~\cite{numrec} and paid attention 
to the vanishing small positive imaginary part of $k^2$.

\begin{figure}
\includegraphics[scale=0.7]{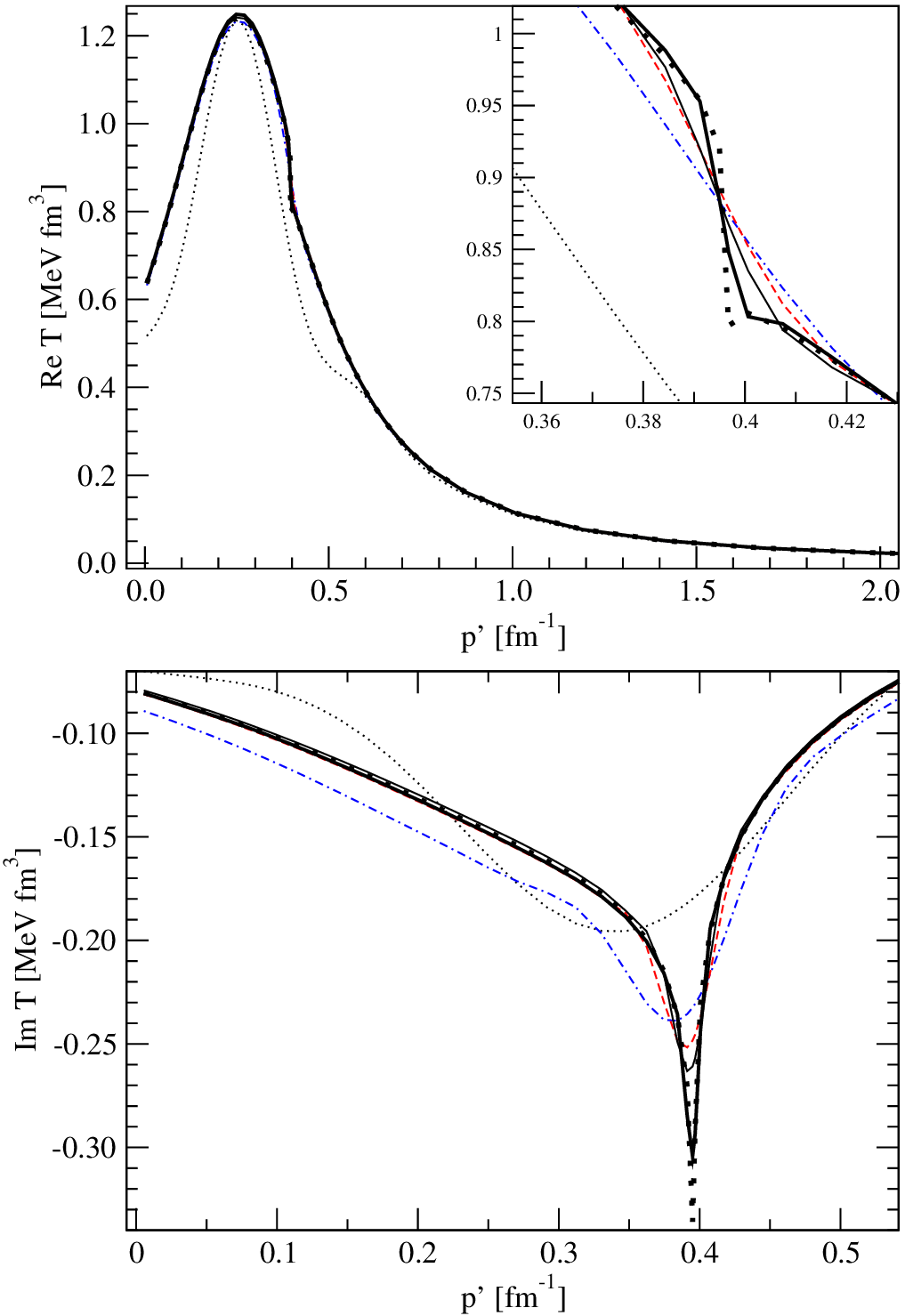}
\caption{(color online) Limiting behaviour of the real (upper) and the
imaginary (lower) parts  of the
off-the-energy-shell screened t-matrix $t_c^R(p,p',x)$ at
$E_p^{lab}=13$~MeV, $p=0.36$~fm$^{-1}$, and $x=0.71$ as a function of the
$p'$
momentum for $n=4$ and different values of the screening radius
$R$: $R=20$~fm (dotted line), $R=60$~fm (dashed-dotted line),
$R=120$~fm (dashed line), $R=180$~fm (thin solid line), $R=500$~fm
(thick solid line). The pure Coulomb off-shell result of
Eq.(\ref{eq.7a}) is given by thick dots. The half-shell situation is reached
at $p'= k = \sqrt{m_p E_{c.m.}}=0.396$~fm$^{-1}$. In the insert a discontinuity
develops if $p'$ approaches $k$ from below or above (see text).}
\label{fig7}
\end{figure}

As an illustration of our general results we show in Fig.~\ref{fig7} the
limiting behavior for the real and imaginary parts of the off-shell screened
Coulomb
t-matrix $ t_c^ R( p,p',x) $
at  $E_p^{lab} = 13$~MeV and fixed $ p $ and $ x $ values as a function of
$p'$. This energy corresponds to the on-shell momentum $k=0.396$~fm$^{-1}$.
 The small $R$-values $R= 20$~fm and $R=60$~fm  are quite insufficient,
especially for the imaginary
part, to reach the pure Coulomb off-shell values. For the higher $R$-values,
$R=120$~fm and beyond, $t_c^ R( p,p',x)$ converges very well and the limit
coincides, as expected, with the pure Coulomb off-shell t-matrix.
 For $p' = k = 0.396$~fm$^{-1}$ one reaches the half-shell point
 and in  its  neighborhood
 a discontinuity develops with increasing $ R $-values, if one
approaches $k$ from below or
above. That discontinuity is  very well known to exist for the pure
half-shell
  Coulomb t-matrix and reproduced for the convenience of the reader:
\begin{equation} 
 \label{eq16}
t_c( p,p',x,E = \frac{k^ 2}{m_p})  =   \frac{e^ 2}{2 \pi^ 2} \frac{1}{|
\vec p - \vec p~'|^ 2}
 \frac{2 \pi \eta} { 1 - e^ {-2 \pi \eta}}
  e^ {i \eta \ln {\frac{ k^ 2 - p^ 2}{ 4 k^2 | \vec p - \vec p~'|^ 2}}}
 e^ { i \eta \ln | k^ 2 - p'^ 2| }
\begin{cases}
 1 & k^2 > p'^2 \cr
   e^{-\pi\eta}  & k^2 < p'^2 \cr
\end{cases}
\end{equation}

\begin{figure}
\includegraphics[scale=0.7]{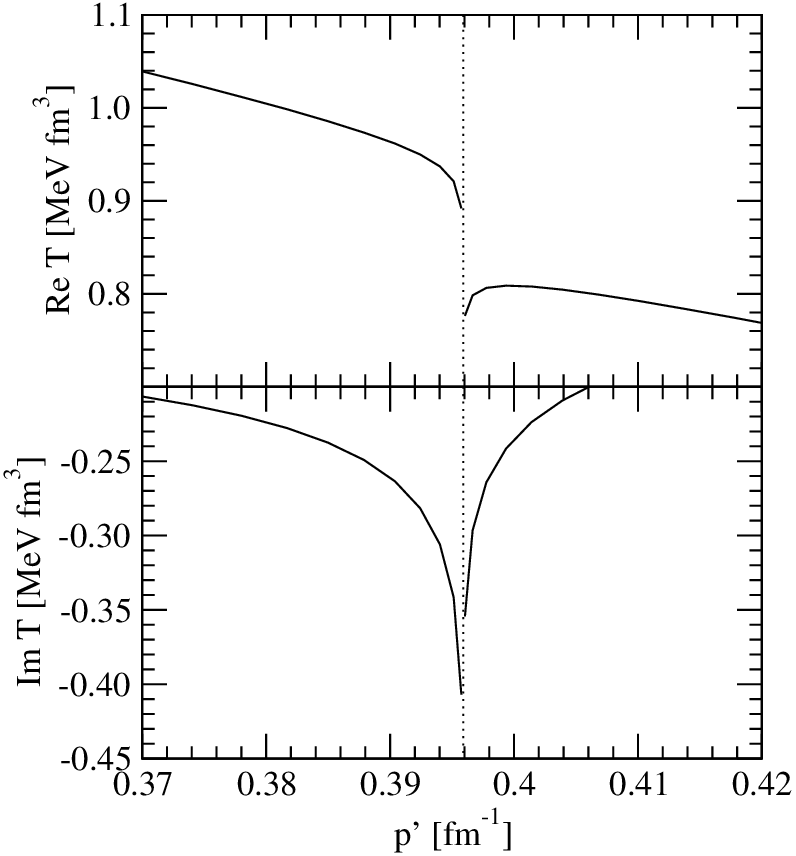}
\caption{
The discontinuity of the real (upper row) and the
imaginary (lower row) parts of the
off-the-energy-shell pure Coulomb t-matrix $t_c(p,p',x)$ (given by Eq.(\ref{eq16})) 
near $p'= k = \sqrt{m_p E_{c.m.}}=0.396$~fm$^{-1}$
at
$E_p^{lab}=13$~MeV, $p=0.361$~fm$^{-1}$, and $x=0.71$ as a function of the
$p'$ momentum.}
\label{fig7a}
\end{figure}

That discontinuity is separately shown in Fig.~\ref{fig7a} for both parts of
the pure Coulomb t-matrix, for predictions based on Eqs~\ref{eq.7}-\ref{eq.7a}.
The screened
Coulomb t-matrix is, of course, continuous at $ p' = k$ for each fixed $
R$-value, but nevertheless
 the tendency to develop that discontinuity can be seen with increasing  R.

  Now as has been emphasized in \cite{ford1964,ford1966} that discontinuity
would be absent if the limits $ p' \rightarrow
k$ and $ R \rightarrow \infty$ are performed such that $ | p'-k| R
\rightarrow 0$.  To show that numerically a  more subtle  investigation is
required, which we did not
undertake.

\begin{figure}
\includegraphics[scale=0.8]{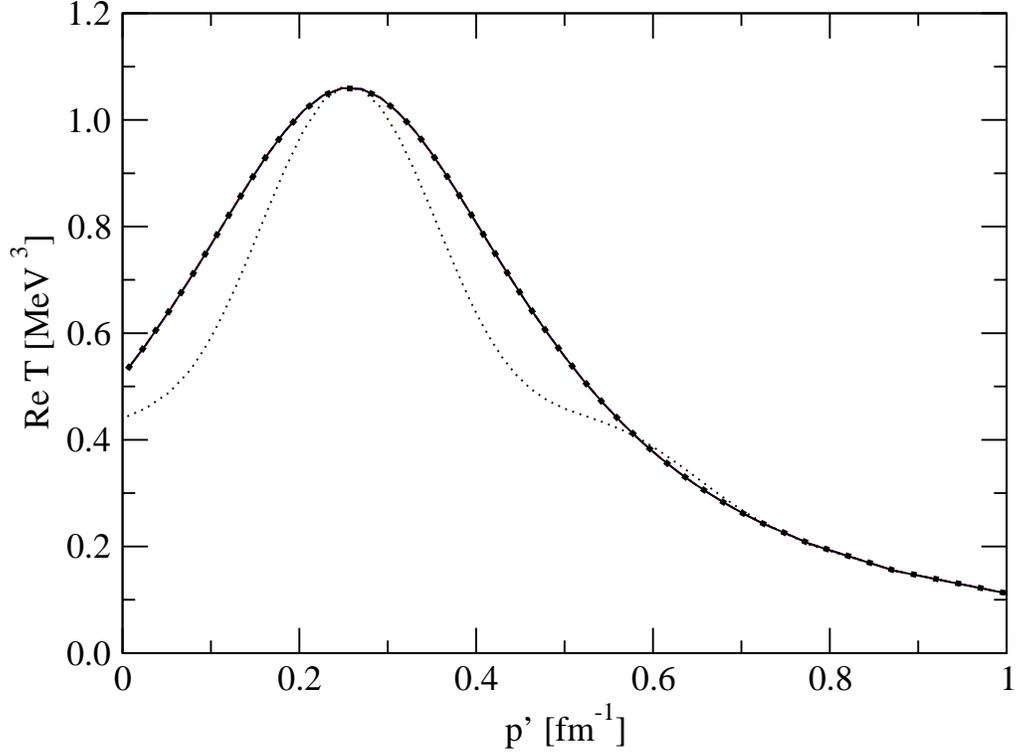}
\caption{(color online) Limiting behaviour of the real part  of the
off-the-energy-shell screened t-matrix $t_c^R(p,p',x)$ at negative energy
$E=-13$~MeV, $p=0.36$~fm$^{-1}$, and $x=0.71$, as a function of the $p'$
momentum for $n=4$ and different values of the screening radius
$R$: $R=20$~fm (dotted line), $R=60$~fm (dashed-dotted line),
$R=120$~fm (dashed line), and $R=180$~fm (solid line).
 The pure Coulomb off-shell result of
Eq.(\ref{eq.7a}) is given by thick dots. } \label{fig8}
\end{figure}

Fig.~\ref{fig8} exemplifies the situation for the off-shell t-matrix at
negative energy $E_p^{lab}= - 13$~MeV, where the t-matrix is real. There the
limit is
reached already around $R= 20$~fm.

\begin{figure}
\includegraphics[scale=0.7]{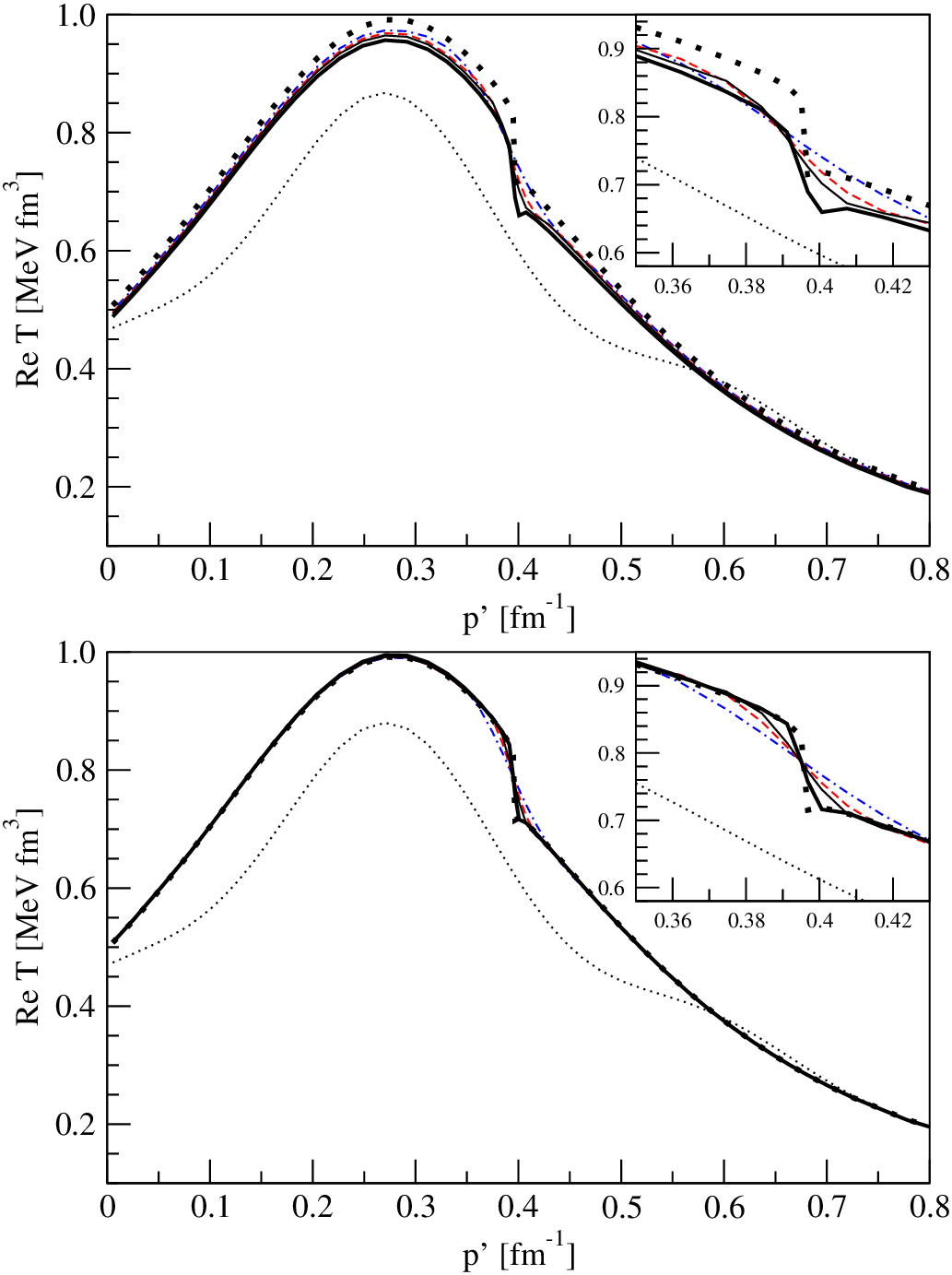}
\caption{(color online) The real part of the half-the-energy-shell
screened t-matrix $t_c^R(p,p,x)$ before (upper row) and after (lower row)
the renormalization. The proton lab. energy is
$E=13$~MeV, and $x=0.71$. The screening potential was taken with $n=4$ 
and different values of the screening radius
$R$: $R=20$~fm (dotted line), $R=60$~fm (dashed-dotted line),
$R=120$~fm (dashed line), $R=180$~fm (solid line)
and $R=500$~fm (thick solid line)
 The pure Coulomb half-shell result of
Eq.(\ref{eq8}) is given by thick dots. } \label{fig7b}
\end{figure}

\begin{figure}
\includegraphics[scale=0.7]{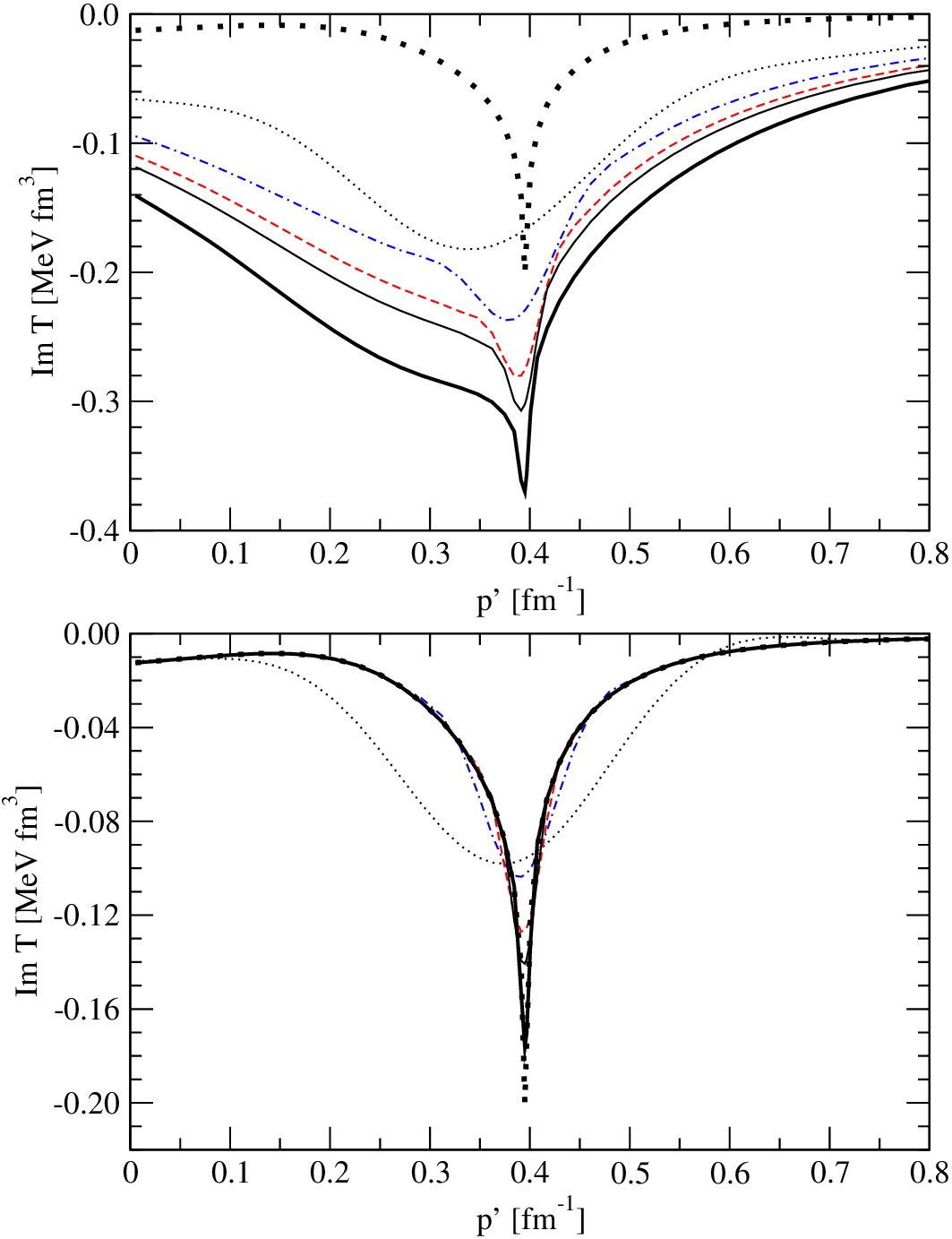}
\caption{(color online) The same as in Fig.~\ref{fig7b} but
for imaginary part of he half-the-energy-shell t-matrix.}
\label{fig7c}
\end{figure}

Let us turn now to the half-shell pure Coulomb t-matrix, which is 
analytically given by~\cite{kok81}.
\begin{eqnarray}
< \vec p~' | t_c^R(\frac {k^2} {m_p}) | \vec k > \rightarrow
C_0 e^{i\sigma_0} \frac{k\eta}{\pi^2 q^2} 
(\frac {p'^2-k^2}{q^2})^{i\eta}\;,\label{eq8}
\end{eqnarray}
where $\vec q = \vec{p\;'} - \vec k$ is the momentum transfer, $\sigma_0=\Gamma(1+i\eta)$
is the pure Coulomb phase shift and $C_0^2=\frac{2\pi\eta}{\exp^{2\pi\eta}-1}$
is the Coulomb penetrability.
The direct comparison of this limit and the screened Coulomb half-shell 
t-matrix is shown in Figs.~\ref{fig7b} and ~\ref{fig7c} for the real and the imaginary part of $t$
at $E_p^{lab}= 13$~MeV,
respectively. On both figures, in upper row discrepancy due to the
oscillating factor $e^{i\Phi_R(k)}$~\cite{taylor1,taylor2,Alt78}
is seen. After removing that factor (by procedure called renormalization) 
the screened half-shell t-matrix approaches in the limit $R \rightarrow \infty$ the 
half-shell Coulomb t-matrix, what is shown in lower row of Figs.~\ref{fig7b} and ~\ref{fig7c}.
The results of renormalization are striking, especially for the imaginary part of
the half-shell screened t-matrix.
The screening radii about $R=60$ fm is sufficient to describe 
the pure Coulomb t-matrix, however for smaller angles one has to go 
even to higher $R$'s.

Finally let us regard the screened on-the-energy-shell t-matrix element,
which  acquires  for
large R-values  the oscillating factor
$e^{2i\Phi_R(k)}$~\cite{taylor1,taylor2,Alt78}.
After renormalization with this factor 
the screened on-the-energy-shell t-matrix approaches in the limit $R
\rightarrow
\infty$ the Coulomb scattering amplitude $A_C(\theta)$~\cite{Alt78,henrybook}
\begin{eqnarray}
 t_c^ R( k,k,x)|_{renormalized} \equiv e^{-2i\Phi_R(k)} t_c^ R( k,k,x)
\rightarrow& - \frac {2} {m_p} 
\frac {A_C(\theta)} {(2\pi)^2} = \frac {2} {m_p(2\pi)^2} \frac
{m_p e^2} {4k^2} \frac { e^{-i\eta ln(sin^2\frac {\theta} {2}) } }
 {sin^2 \frac {\theta}{2} } ~.
\label{eq.9}
\end{eqnarray}

\begin{figure}
\includegraphics[scale=0.9]{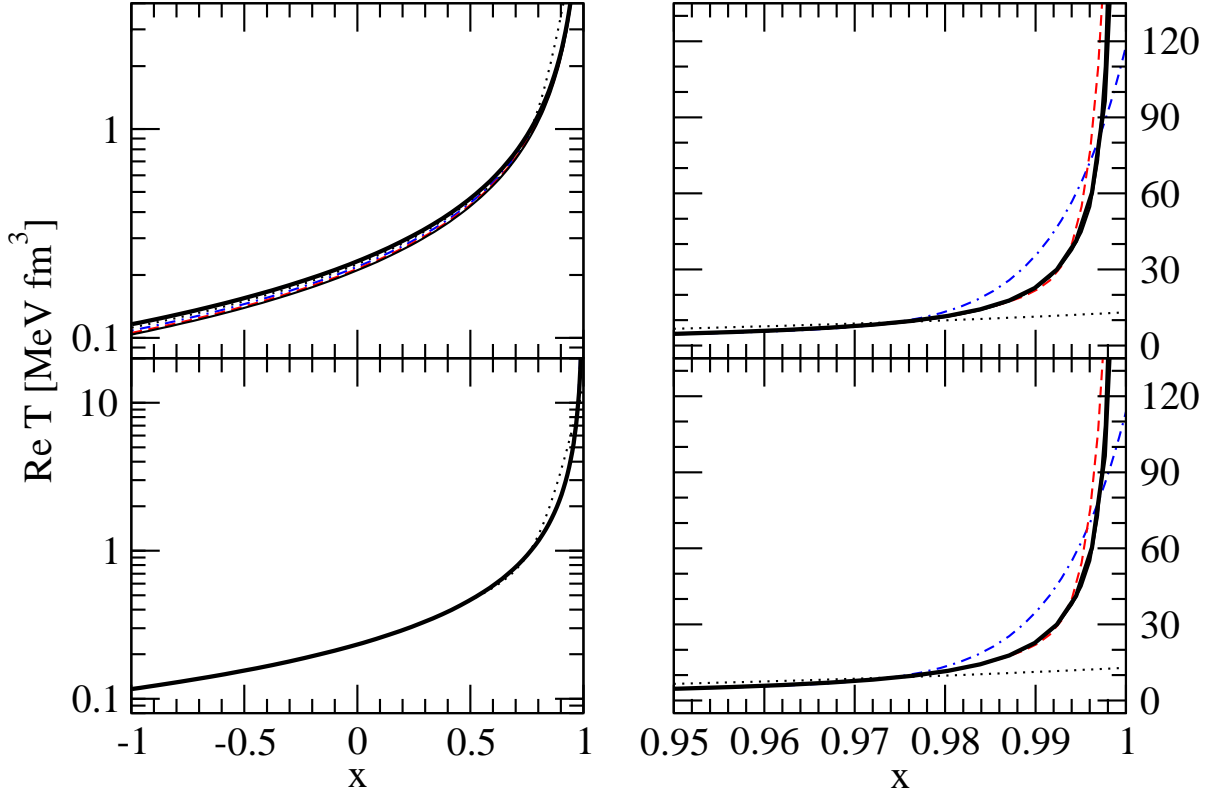}
\caption{(color online) The real part of the on-the-energy-shell
screened t-matrix $t_c^R(p,p,x)$ (upper row) and the corresponding
 renormalized t-matrices (lower row)
 at $E_p^{lab}=13$~MeV as a function of
$x$ for $n=3$ and different values of the screening radius $R$:
$R=20$~fm (dotted line), $R=60$~fm (dashed-dotted line),
$R=120$~fm (dashed line), $R=180$~fm (thin solid line). The
Coulomb on-shell amplitude of Eq.(\ref{eq.9}) is given by thick
solid line. The left and the right column
differs only in the scale of the x-axis.} \label{fig9}
\end{figure}

\begin{figure}
\includegraphics[scale=0.9]{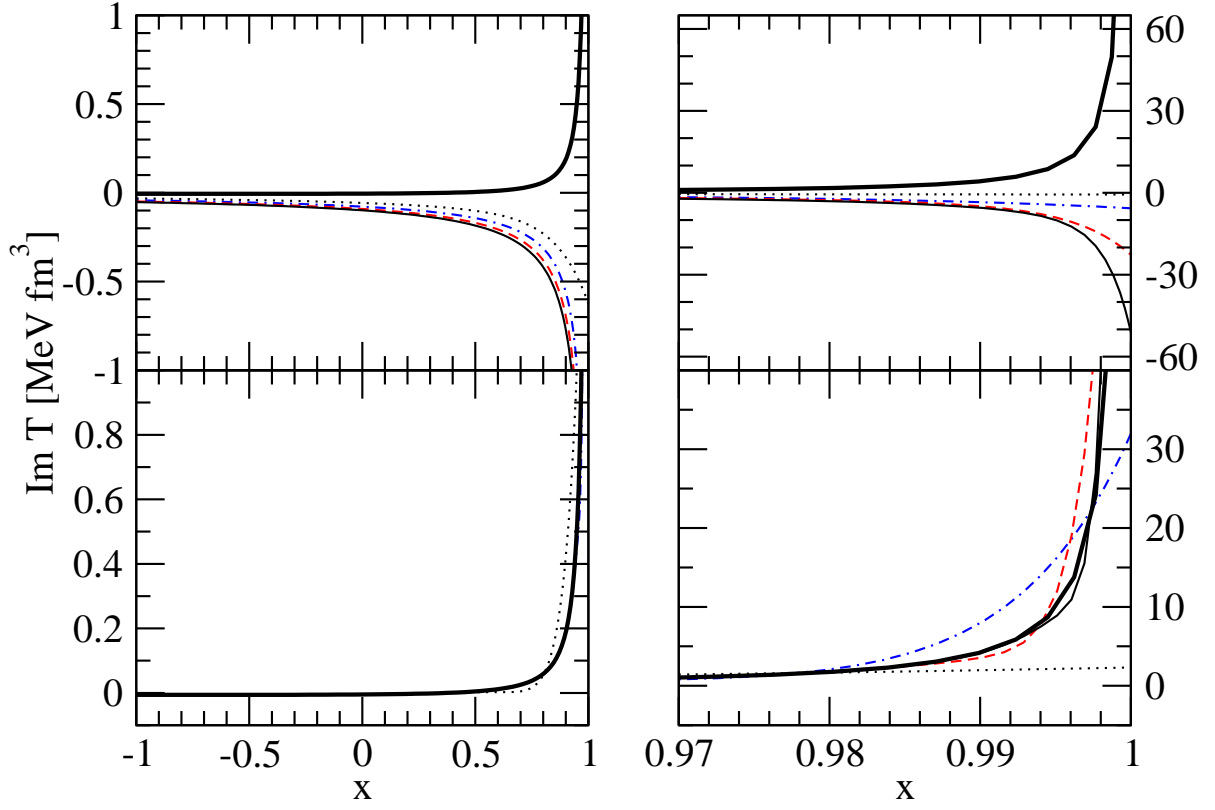}
\caption{(color online) The imaginary part of the on-the-energy-shell
screened t-matrix $t_c^R(p,p,x)$ (upper row) and the corresponding
 renormalized t-matrices (lower row)
 at $E_p^{lab}=13$~MeV as a function of
$x$ for $n=3$ and different values of the screening radius $R$. For the
description of lines see Fig.\ref{fig9}.}
 \label{fig10}
\end{figure}

This is demonstrated in Fig.~\ref{fig9} for the real part and in
Fig.~\ref{fig10} for the imaginary part of the on-shell t-matrix at
$E_p^{lab}= 13$~MeV. The upper rows show unrenormalized screened t-matrices
while the lower rows show them after renormalization. It is clearly seen
that the renormalization is required to get the Coulomb on-shell amplitude
(shown by the thick solid line).
  The effect of  renormalization is very pronounced for the imaginary part.
It is  negative without renormalization  and shows a strong dependence on
the screening radii $R$. After renormalization and for sufficiently large $
R$-values it perfectly overlaps
with the imaginary part of the pure Coulomb on-shell
 amplitude.
 At forward angles, which corresponds to $x$ close to 1, $R=180$~fm is required to reach good agreement with the
pure on-shell
Coulomb t-matrix.

\section{Summary and outlook}
We numerically solved the 3-dimensional LS equation for a general off-shell
screened Coulomb t-matrix with different types of screening.
That t-matrix taken on-shell together with a finite number of partial wave
projected t-matrices
generated by the sum of the screened Coulomb force and the nuclear force and
corrected for the
 partial wave projected screened pure Coulomb t-matrix leads to correct pp
observables at
 suitably chosen finite screening radii. The renormalization phases drop
out automatically in the pp observables
 since they are products of the full t-matrix and its complex conjugate.
 Thus renormalization is not required and even the  knowledge of the
renormalization phase is not required when calculating observables.

Finally we numerically checked that the screened 3-dimensional Coulomb 
t-matrix has the off-shell,
half-shell and on-shell limits in relation to the general unscreened pure
Coulomb t-matrix, 
 which have been analytically predicted in the literature. We felt that this
numerical 3-dimensional 
investigation  supplements well those previous analytical studies, which
lack mathematical rigor
 when summing partial wave results to infinite order.

  The resulting 3-dimensional screened general Coulomb t-matrix will be used
in a forthcoming Faddeev
treatment of pd reactions.

\section*{Acknowledgments}
This work was supported by the 2008-2011 Polish science funds as a
 research project No. N N202 077435. It was also partially supported by the
Helmholtz
Association through funds provided to the virtual institute ``Spin
and strong QCD''(VH-VI-231) and by
  the European Community-Research Infrastructure
Integrating Activity
``Study of Strongly Interacting Matter'' (acronym HadronPhysics2, Grant
Agreement n. 227431)
under the Seventh Framework Programme of EU.
 The numerical calculations were
performed on the supercomputer cluster of the JSC, J\"ulich,
Germany.

\end{document}